%% file: SSaC_VTC_v4_arxiv.tex
\documentclass[conference]{IEEEtran}
 \usepackage{amsmath,amssymb}
 \usepackage{subfigure}
 \usepackage{graphicx,graphics,color,psfrag}
 \usepackage{cite,balance}
 \usepackage{caption}
 \allowdisplaybreaks
 \usepackage{algorithm}
 \usepackage{accents}
 \usepackage{amsthm}
 \usepackage{bm}
 \usepackage{algorithmic}
 \usepackage[english]{babel}
 \usepackage{multirow}

 \usepackage{enumerate}
 \usepackage{cases}
 \usepackage{stfloats}
 \usepackage{dsfont}
 \usepackage{color,soul}
 \usepackage{amsfonts}
 \usepackage{cite,graphicx,amsmath,amssymb}
 \usepackage{subfigure}
 \usepackage{fancyhdr}
 \usepackage{hhline}
 \usepackage{graphicx,graphics}
 \usepackage{array,color}

 \usepackage{amsmath}


\newcounter{parentnumber}

\include{header}

\begin{document}

\title{\Huge {Symbiotic Sensing and Communications Towards 6G: Vision, Applications, and Technology Trends}}
\author{Zhiqin Wang$^\dag$, Kaifeng Han$^\dag$, Jiamo Jiang$^\dag$, Zhiqing Wei$^\ddag$, Guangxu Zhu$^\S$, Zhiyong Feng$^\ddag$,\\Jianmin Lu$^*$ and Chunwei Meng$^\ddag$\\
{$\dag$ China Academy of Information and Communications Technology, Beijing, China }\\{ $\ddag$ Beijing University of Posts and Telecommunications, Beijing, China}\\{ $\S$ Shenzhen Research Institute of Big Data, Shenzhen, China}\\ { $*$ Huawei Technologies Co., Ltd., Shenzhen, China}\\\vspace{-10pt} { Email: zhiqin.wang@caict.ac.cn}}
\maketitle

\begin{abstract}
Driven by the vision of \emph{intelligent connection of everything and digital twin} towards 6G, a myriad of new applications, such as immersive extended reality, autonomous driving, holographic communications, intelligent industrial internet, will emerge in the near future, holding the promise to revolutionize the way we live and work. These trends inspire a novel technical design principle that seamlessly integrates two originally decoupled functionalities, i.e.,  wireless communication and sensing, into one system in a symbiotic way, which is dubbed \emph{symbiotic sensing and communications} (SSaC), to endow the wireless network with the capability to ``see" and ``talk" to the physical world simultaneously. Noting that the term SSaC is used instead of ISAC (integrated sensing and communications) because the word ``symbiotic/symbiosis" is more inclusive and can better accommodate different integration levels and evolution stages of sensing and communications. Aligned with this understanding, this article makes the first attempts to clarify the concept of SSaC, illustrate its vision, envision the three-stage evolution roadmap, namely neutralism, commensalism, and mutualism of SaC. Then, three categories of applications of SSaC are introduced, followed by detailed description of typical use cases in each category. Finally, we summarize the major performance metrics and key enabling technologies for SSaC.
\end{abstract}


\section{Introduction}
With the commercialization of 5G, both academia and industry have shifted their focuses toward the development of 6G solutions for 2030 and beyond. It is envisioned that diversified new killer applications and services, such as immersive multisensory \emph{extended reality} (XR), autonomous driving, holographic communications, intelligent healthcare, intelligent industrial, digital twins, etc., will appear and be supported by 6G networks \cite{imt2021whitepaper6G}. To meet the ever-increasing demands, the 6G will evolve from a pure communication infrastructure toward a comprehensive platform accommodating functionalities such as wireless communication, sensing, computing, and \emph{artificial intelligence} (AI), and utilize all of them to revolutionize the way we live and work. This ongoing trend, together with the demand for higher spectrum utilization, has first driven the integration of two important functionalities, i.e., communications and sensing, into a single platform, where they work together in a symbiotic way to pursue mutual benefits, leading to an emerging research area named the \emph{symbiotic sensing and communications} (SSaC). Endowed by SSaC, the future wireless network could \textbf{see} (sense) and \textbf{talk} (communicate) to the physical world simultaneously in a cost-effective way.

It is worth noting that different similar terms, such as \emph{integrated sensing and communications} (ISAC) \cite{Tong20216G}, \cite{tan2021ISAC}, \emph{joint radar and communication} (JRC) \cite{feng2020jsac}, \emph{dual-functional radar and communications} (DFRC) \cite{hassanien2019dfrc}, \emph{joint communication and radar/radio sensing} (JCAS) \cite{zhang2021enablingjcrs}, have been used in the literature. In this article, we advocate the use of the term SSaC because the word ``\textbf{symbiotic/symbiosis}" is more inclusive and can better accommodate different integration levels and evolution stages of \emph{sensing and communications} (SaC).

Recently, SSaC has attracted substantial research interests in both academic and industry. There has been an increasing number of survey or tutorial papers on this topic. For example, the concepts, characteristics, applications, and the latest research progress of JRC are introduced in \cite{feng2020jsac}. In \cite{zhang2021enablingjcrs}, a comprehensive survey on JCAS techniques is provided and a novel network architecture, named \emph{perceptive mobile network} (PMN), is proposed to show the feasibility that integrates radio sensing into the existing cellular mobile network. Authors in \cite{liu2021fundamentalofisac} surveyed the recent studies on fundamental limits of ISAC, in which the major performance metrics are summarized, providing useful guideline for the practical network design. In \cite{cui2021isacIoT}, various use cases in ISAC-enabled IoT networks are presented. More examples of ISAC applications are categorized and elaborated in \cite{Tong20216G}, where the research directions for ISAC air interface are also outlined.

Despite these scattered previous research efforts on SSaC, there still lacks of an unified understanding of SSaC in terms of its concept, vision and evolution roadmap. To close the gap, in this article, we make the first attempt to present our understanding of SSaC in a unified viewpoint, with focus on presenting concept, vision, evolution roadmap, application scenarios, and supporting use cases as well as the key performance metrics and enabling techniques.

\section{Concept and Vision of Symbiotic Sensing and Communications }
In this section, we first clarify the concepts of SSaC, and then present the evolution roadmap of SSaC towards 6G. It is believed that SSaC will finally become one of the main services in the future 6G.

\begin{figure}[t]
\centering
\includegraphics[width=9cm]{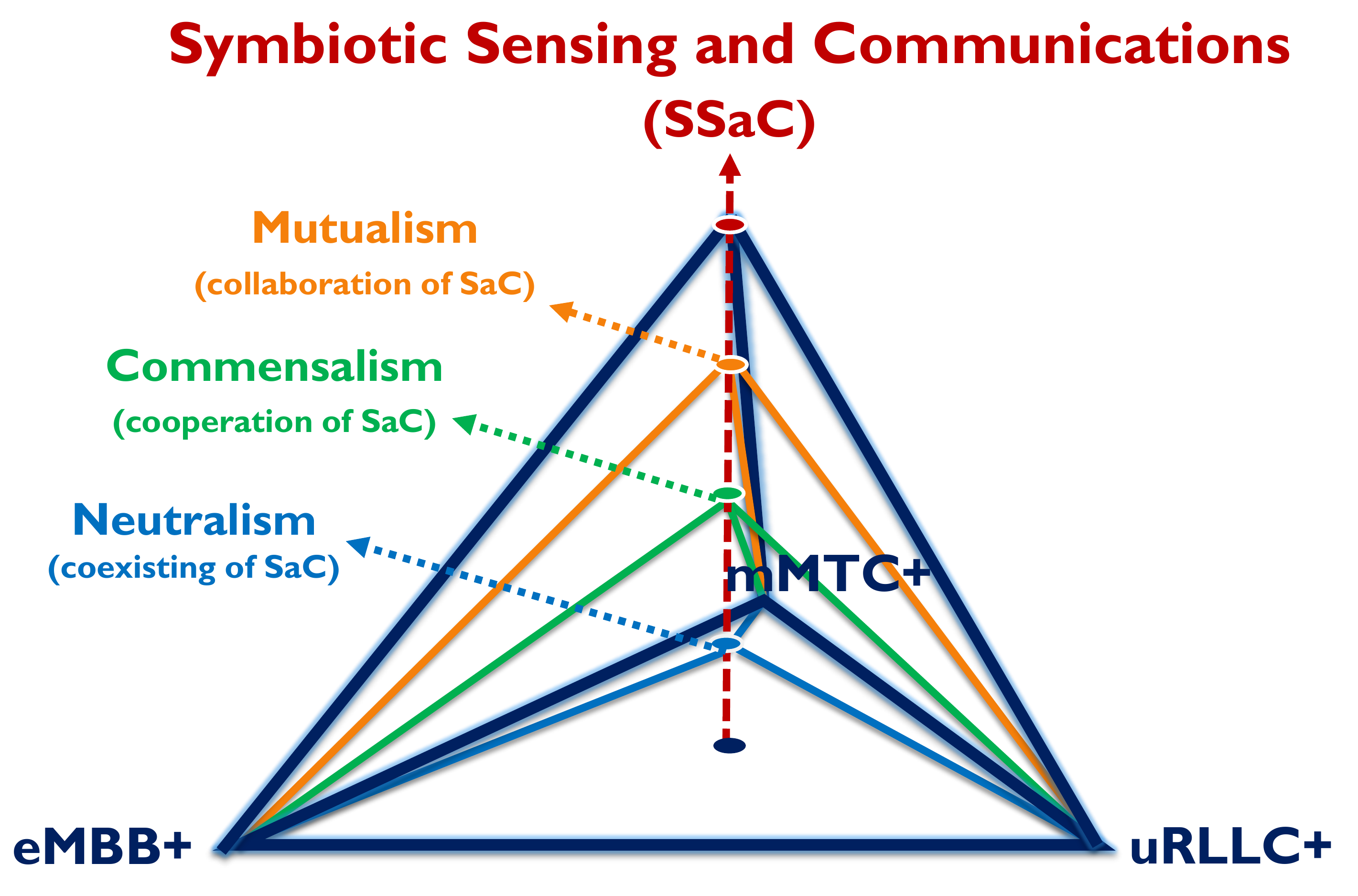}
\caption{Evolution of SSaC, including neutralism, commensalism, and mutualism of SSaC. Finally, SSaC will be a main service in 6G.}\label{Fig:EvolutionOfSSaC}
\vspace{-10pt}
\end{figure}

\subsection{Concept and Features of SSaC}
As one of the new design paradigms towards 6G, SSaC aims to effectively utilize the capabilities and mutual benefits of both communication and sensing to provide diverse services, ranging from immersive applications (e.g., XR) to intelligent applications (e.g., autonomous driving). To understand the concept of SSaC, several points are clarified as listed below.
\subsubsection{The scope of sensing in SSaC}
The conventional concept of sensing can be implemented in two different ways, namely the \textbf{sensing via electromagnetic wave signals} [e.g., \emph{radio frequency} (RF) signals\footnote{RF signals include sub-6 GHz/mmWave/THz/WiFi/UWB signals, etc.}, optical signals\footnote{Optical signals includes visible-light/infrared signals, etc.}, etc.], and \textbf{sensing via sensors or sensor systems} (e.g., camera, microphone, temperature and humidity sensor, etc.). The former emphasizes the use of wireless signals for sensing, while the latter emphasizes using sensor networks to obtain sensory information. The concept of sensing in SSaC is more related to the former, focusing on leveraging the electromagnetic wave signals, especially the RF signals for sensing, which facilitates the integration of wireless sensing capability into the communication networks.
\subsubsection{The meaning of symbiosis}
The word ``symbiosis" comes from the concept in ecology, referring to biological interactions between two different organisms. This term is more inclusive and better accommodate the different degrees of integration (or different symbiotic mechanisms) between sensing and communications, which are further discussed in Section II-B.
\subsubsection{The concept of SSaC}
The main design principle in SSaC is to seamlessly integrate and effectively utilize both the wireless sensing and communications in the same system (by sharing the same frequency, signaling, hardware, etc.) in a reciprocal and symbiotic way. Specifically, SSaC could utilize the electromagnetic wave signals to wirelessly detect/localize/track/image the objects,  recognize different activities or status, or even reconstruct the environments, and the sensing results can be used to enhance the performance of radio access and resource management in wireless communications. Furthermore, SSaC is able to support diverse applications, including location and trajectory based applications, status recognition based applications, and environmental reconstruction based applications, providing high quality services and user experience for 6G.

\subsection{Evolution Roadmap and Vision of SSaC}
As the integration level of SaC continues to increase, different stages of symbiosis, including neutralism, commensalism, and mutualism, will exist and together form the evolution roadmap of SSaC in 6G era, as shown in Fig. \ref{Fig:EvolutionOfSSaC}.

\emph{Stage 1: Neutralism of SaC}, namely, the co-existence of SaC, meaning that sensing and  communications may use the separated signals and processing procedures, but share the same resources (e.g., spectrum frequency, antenna array, etc.) to achieve higher spectral and hardware efficiencies. The key research issues in this stage could be designing the efficient interference cancellation and management techniques, so SaC can operate without unduly interfering with each other \cite{zhang2021enablingjcrs}.

\emph{Stage 2: Commensalism of SaC}, also called the cooperation of SaC, referring to the level that sensing and communications can operate using the same hardware and signaling (e.g., sharing a waveform), and exploit the joint knowledge to improve the performance of one system without affecting the other, i.e., sensing-assisted communication design, or communication-assisted sensing design. The research directions include joint design ranging from waveform, coding schemes, to signal processing algorithms.

\emph{Stage 3: Mutualism of SaC}, namely, the collaboration of SaC, means that sensing and communications are completely coordinated and collaborated in all dimensions including spectrum, hardware, signaling, protocol, networking, etc.  They could be mutually promoted and benefited, which is the highest-level symbiosis between SaC. In this stage, both SaC will obtain the highest performance gain simultaneously to support a wide range of  new applications and services.

To conclude, in 6G era, SSaC will grow to be one of the main services in addition to eMBB, uRLLC, and mMTC services supported by 5G. As shown in Fig. \ref{Fig:EvolutionOfSSaC}, SSaC will coexist with \textbf{eMBB+}, \textbf{uRLLC+}, \textbf{mMTC+}, forming a ``\textbf{tetrahedron}" that contains the fundamental capabilities and services for 6G.

\section{Application Scenarios and Performance Metrics of SSaC}
\begin{figure}[t]
\centering
\includegraphics[width=9cm]{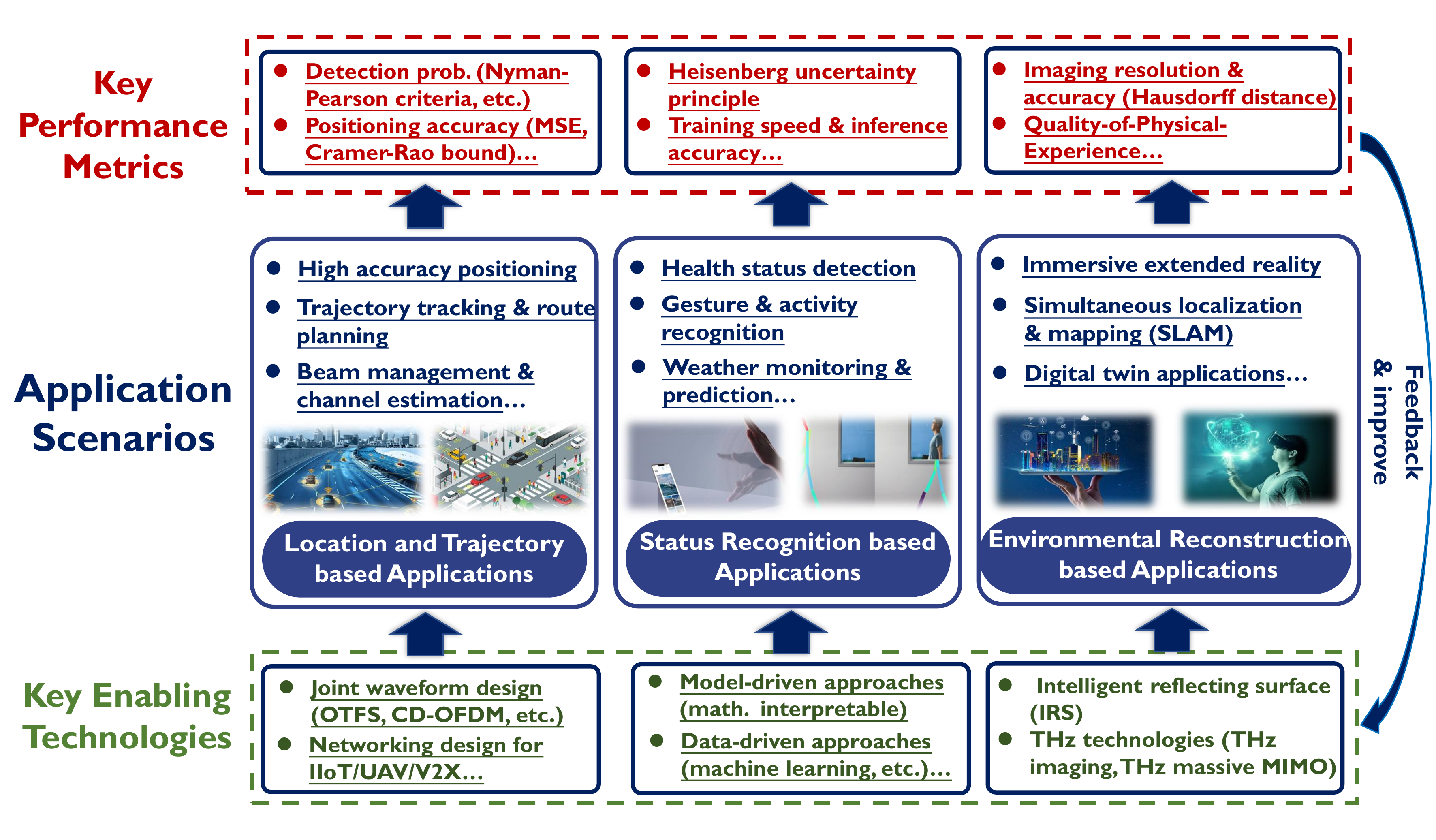}
\caption{Application scenarios, key enabling technologies and performance metrics of SSaC. The performance evolution could be fed back to the technologies design for further improvement.}\label{Fig:ApplicationOfSSaC}
\vspace{-10pt}
\end{figure}

SSaC (including all three stages) is expected to support diverse new applications. In this section, we introduce three typical application scenarios of SSaC and illustrate the most representative use cases for each of the scenario. Then, the corresponding performance metrics are discussed (see Fig. \ref{Fig:ApplicationOfSSaC}).

\subsection{Application Scenarios and Use Cases of SSaC}
This subsection introduces three application scenarios of SSaC, and illustrates several typical use cases for each of them.

\subsubsection{Location and trajectory based applications}
As the essential features for many emerging applications, the high-accurate location and trajectory of targets will be the critical information to be sensed by a SSaC system which the current 5G fails to provide. The typical use cases may include below.

\begin{itemize}
  \item \emph{High accuracy positioning}: SSaC needs to deliver centimeter-level localization accuracy to enable, for example, intelligent navigation for indoor shopping or taxi ride, positioning services for emergency rescue, etc.
  \item \emph{Trajectory tracking and route planning}: The trajectory information of multiple objects (such as drones, vehicles, and other robots) can be used to enable accurate route planning, large-scale neighbor discovery, collision avoidance, vehicle platooning, which help to alleviate traffic congestion and improve road safety.
  \item \emph{Beam management and channel estimation}: By using the location of the target sensed by SSaC, fast and efficient beam sweeping, pairing, and alignment procedures can be achieved, especially for the high-mobility scenarios \cite{feng2020jsac}. Also, the angle and timing information (i.e., AoA/AoD/ToA) obtained via positioning can be further used for communication channel estimation.
\end{itemize}

\subsubsection{Status recognition based applications}
In this scenario, different kinds of status information, including gesture/activities/health status/weather condition,  could be extracted and recognized by jointly using the channel state information (CSI) and machine learning/AI algorithms.

\begin{itemize}
  \item \emph{Gesture and activity recognition}: By analyzing the spectrogram or imaging results obtained through communication channel (e.g., mmWave/THz/WiFi communications), various human motions (such as standing/walking/running) can be accurately recognized. Moreover, the mobile phone could detect the human's gesture via its radio signals or embedded sensors \cite{cui2021isacIoT}.
  \item \emph{Health status detection}: The human breath rate and breath status can be detected and tracked via exploiting the phase variation of CSI. SSaC can also be applied in medical diagnosis and vital signs monitoring by analyzing the RF signals and CSI or using wearable devices \cite{tan2021ISAC}.
  \item \emph{Weather monitoring and prediction}: Since some weather conditions, e.g., rainfall, hailing, could significantly affect the path loss or Doppler  of mmWave links, it is possible to use the cellular network for weather monitoring and prediction, as well as atmospheric observation.
\end{itemize}

\subsubsection{Environmental reconstruction based applications}
One important functionality of SSaC is to sense the important targets or events, and reconstruct the environment with ultra-high-resolution RF images \cite{imt2021whitepaper6G}. The typical use cases include:

\begin{itemize}
  \item \emph{Simultaneous localization and mapping} (SLAM): The main purpose of SLAM is to construct and update the map of an unknown environment. Performing SLAM via RF signals [e.g., mmWave/terahertz(THz)] instead of cameras or laser measurements will be a promising way due to its larger sensing range and relatively low complexity and data amount for processing.
  \item \emph{Immersive extended reality} (XR)\footnote{XR contains \emph{Virtual Reality} (VR), \emph{Augmented Reality} (AR), and \emph{Mixed Reality} (MR).}: To enrich the immersive user experience in XR, SSaC will be used to capture the detailed physical environment and human movement to further increase the fidelity of virtual worlds, supporting immmersive applications such as telepresence.
  \item \emph{Digital twin applications}: By jointly leveraging the SSaC, AI, and sensor networks, it will be feasible to replicate the physical objects in a virtual world via digital replica, called the digital twin. In the digital twin, both historical and real-time data are used to simulate, verify, predict, and control physical objects or processes, offering the optimal solutions to the issues in real physical world \cite{imt2021whitepaper6G}.
\end{itemize}

\subsection{Performance Metrics and Requirements of SSaC}
In this subsection, the key performance metrics for characterizing the fundamental limits of SSaC are proposed.

\subsubsection{Performance metrics for location and trajectory based applications}
The target detection and positioning (i.e., tracking) are two fundamental functions in location and trajectory based applications, which could be characterized by using the metrics discussed below.
\begin{itemize}
  \item \emph{Detection probability}: As one of the basic functions in the SSaC system, target detection refers to determining whether a target is present or absent via analyzing the reflected echo. However, the echo signal is always interfered with by additive noise or clutter, leading to false alarm or miss detection. The mathematical model to measure the probability of false alarms or miss detection is hypothesis testing. Let hypothesizes $H_0$ and $H_1$ denote the absent of target and existence of target, respectively. The false alarm probability is defined by $P_{\mathrm{fal}} = \Pr(H_1|H_0)$ and the detection probability is defined by $P_{\mathrm{det}} = 1 - \Pr(H_0|H_1)$. Given the above parameters, the Neyman-Pearson criterion is used to maximize the detection probability $P_\mathrm{det}$ with a constraint of $P_\mathrm{fal}$. Moreover, \emph{receiver operating characteristic} (ROC) curve is also used to characterize the relation between $P_{\mathrm{fal}}$ and $P_\mathrm{det}$ with given parameters such as SNR, etc.

  \item \emph{Positioning accuracy}: The positioning accuracy highly depends on several parameter estimation results including angle/range/Doppler/velocity. The estimation performance can be characterized by \emph{mean squared error} (MSE) between the estimated parameter and its true value. To allow a tractable analysis on MSE, the \emph{Cramer-Rao bound} (CRB) is widely used as the optimal performance metric that provides a lower bound of the variances of all unbiased estimators \cite{liu2021JRC1}.
\end{itemize}

\subsubsection{Performance metrics for status recognition based applications}
The basic workflow in a status recognition based application mainly consists of three phases, i.e., the sensing signal reception, measurement pre-processing, and recognition, as illustrated in Fig. \ref{Fig:SensingRecogo}. Different performance metrics will thus be used in the course of sensing:
\begin{figure}[t]
\centering
\includegraphics[width=9.3cm]{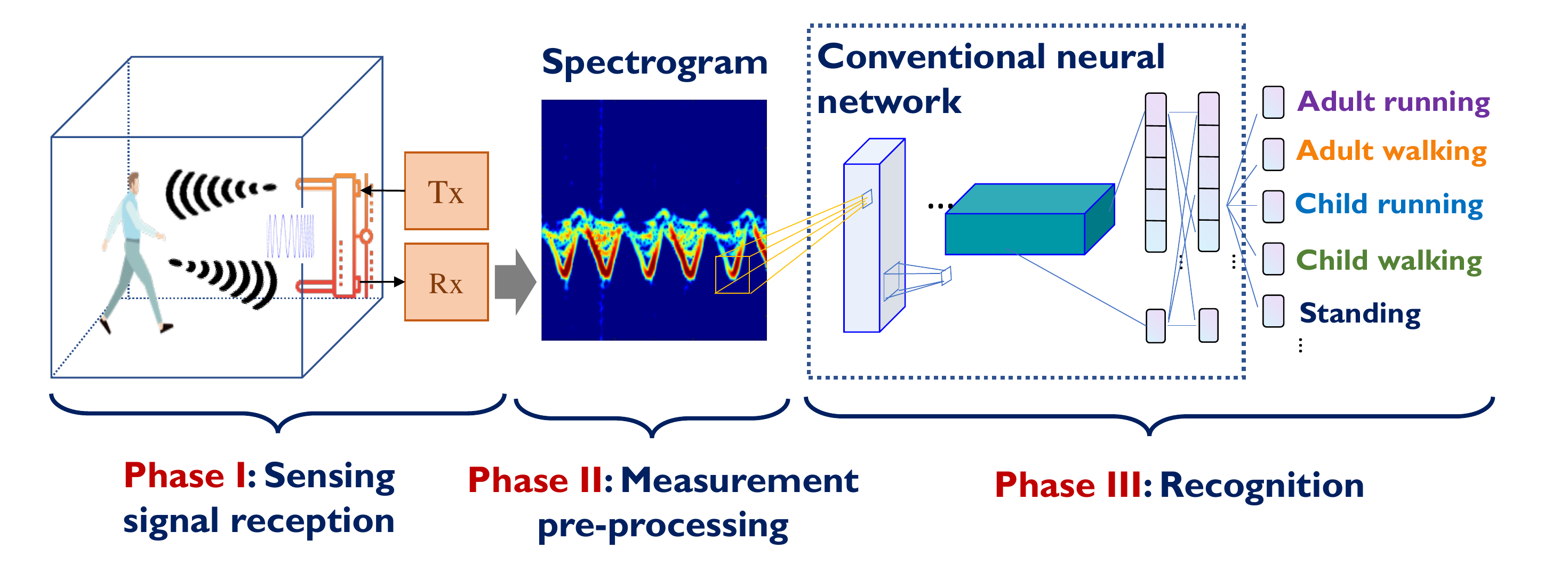}
\caption{Wireless sensing procedures for human motion recognition.}\label{Fig:SensingRecogo}
\vspace{-10pt}
\end{figure}

\begin{itemize}
  \item \emph{Signal-to-interference-and-noise-ratio} (SINR): During the signal reception phase, sensing receivers first receive the wireless signals reflected by sensing objectives. The performance metric used for evaluating the quality of the raw received signals is the SINR.
  \item \emph{Heisenberg uncertainty principle}: Within the measurement pre-processing phase, the raw received signals will go through the \emph{short-time Fourier Transform} (STFT) to yield the spectrogram, which can better expose how the Doppler shift due to motion changes over time. By examining the spectrogram one can thus extract the motion status of the target objective. Related performance metrics in this phase are the time and frequency resolution of the spectrogram. There exists a tradeoff between the two metrics, which is controlled by the width of the time window selected for STFT. Particularly, the increase of time resolution will be at a cost of compromised frequency resolution and vice versa. The tradeoff is known as the Heisenberg uncertainty principle.
  \item \emph{Training speed and inference accuracy}: During the recognition phase, the preprocessed measurements will then fed into a machine learning model, e.g., conventional neural network, for recognition model training and inference. The two key performance metrics used here are training speed and inference accuracy.
\end{itemize}

\subsubsection{Performance metrics for environmental reconstruction based applications}
Many future intelligent and immersive applications like VAR-based gaming, holographic rendering, and digital twins are expected to demand ultra-accurate imaging and ultimate user experience with higher bit-rate and lower latency. The following metrics are essential to evaluate the imaging performance and user experience.

\begin{itemize}
  \item \emph{Imaging accuracy}: As one of popular metrics in imaging applications, the Hausdorff distance is used to quantify the similarity of two images and then evaluate the image retrieval performance \cite{zhang2021vehicularsensing}. Specifically, consider two images\footnote{The image is composed discrete points.} $\mathcal{A}$ (real one) and $\mathcal{B}$ (retrieved one), the corresponding Hausdorff distance is given by
      \begin{align}\label{Eq:HausdorffDistance}
        \mathcal{H}(\mathcal{A}, \mathcal{B}) = \max(\mathrm{h}(\mathcal{A}, \mathcal{B}), \mathrm{h}(\mathcal{B}, \mathcal{A})), \nn
      \end{align}
      where $\mathrm{h}(\mathcal{A}, \mathcal{B}) = \max_{a\in \mathcal{A}} \min_{b \in \mathcal{B}}\parallel a - b \parallel$.

  \item \emph{Quality-of-Physical-Experience} (QoPE): To measure the user experience of human-centric type services in SSaC (e.g., cloud-XR, holographic communications), the QoPE is designed and proposed that integrates both raw wireless metrics and human physical factors \cite{saad2020vision6G}. In other words, QoPE could not only characterize the service-related hardware and software performance, but also quantify the real feeling or experience of human user.
\end{itemize}

\section{Enabling Technologies and Challenges in SSaC}
Recently, extensive research efforts have been dedicated to developing enabling technologies, ranging from waveform design, transceiver architecture, to signal processing and machine learning algorithms, for supporting SSaC. In this section, we point out several technology trends in SSaC.

\subsection{Enabling Technologies for  Location and Trajectory based Application}
Joint waveform design and networking design are two cornerstone technologies for positioning and tracking in SSaC. This is because the optimal waveform help to improve the performance of both target detection and data transmission, while the networking design enables large-scale sensing for multiple targets.

\begin{itemize}
  \item \emph{Joint waveform design}: One key research problem in SSaC is the optimal design of signal waveform for joint SaC. Lots of studies have investigated this issue. Specifically, \emph{orthogonal frequency division multiplex} (OFDM) waveform is attracting much attention in SSaC design due to its flexible subcarrier modulation and high spectrum utilization. To further improve both SaC capabilities and suppress their mutual interference, a novel \emph{code-division OFDM} (CD-OFDM) waveform and corresponding signal processing algorithms are proposed in \cite{chen2021CD-OFDM} that exploit the code-division gain (or spreading gain) for multi-user SSaC systems. In addition, another promising candidate is the \emph{orthogonal time frequency space} (OTFS) based waveform \cite{yuan2021OTFS}, which could support reliable SaC functions in high-mobility scenarios, e.g., vehicular networks.

  \item \emph{Networking design}: With the increasing popularity of intelligent industry and smart cities, a large number of wireless devices will be densely deployed in every corner of our life. Therefore, SSaC should be capable of supporting a massive number of devices simultaneously. To this end, the networking design for SSaC needs to be reckoned with. One promising way is to design a distributed and cooperative SaC network by optimally grouping and scheduling devices, e.g., realising the networked sensing under a cellular topology \cite{zhang2021enablingjcrs}.
\end{itemize}

\subsection{Enabling Technologies for Status Recognition based Application}
As two categories of key techniques in status recognition, model-driven approaches and data-driven approaches have great potential in extracting key features from sensed signals to offer motion detection or gesture recognition with high classification accuracy \cite{wang2018wirelesssensing}.

\begin{itemize}
  \item \emph{Model-driven approaches}: Model-driven approaches refer to those use geometry or statistical relationship between nodes or observations to make the status recognition in a mathematically interpretable manner. The advantage of model-based approaches is the low complexity that makes time-consuming training procedure not required. However, it is generally challenging if not impossible to build an accurate geometry or statistical model to characterize the sensing environment.
  \item \emph{Data-driven approaches}: For data driven approaches, machine learning algorithms are used to directly learn the mapping between input features and output prediction in a black box manner. The advantage of the data-driven approaches is that they do not rely on the accurate modeling of the sensing environment. However, how to acquire a sufficient number of labeled data for learning model training is the key challenge to be reckoned with.
\end{itemize}

\subsection{Enabling Technologies for Environmental Reconstruction based Applications}
Recent breakthroughs in the fabrication of intelligent reflecting surface  and THz techniques  have made possible the finely sensing of the various objects, based on which the accurate environmental reconstruction can be achieved.

\begin{itemize}
  \item \emph{Intelligent reflecting surface} (IRS): Currently, an innovative concept named the smart radio environment has been proposed with the vision of making the wireless channels fully controllable (e.g., creating LoS links, reshaping channel realizations, improving channel rank, etc.) by leveraging the low-cost IRS techniques. Beyond communication purposes, IRS can also be used to deliver the accurate and efficient radio-based environment sensing and imaging in both indoor and outdoor scenarios \cite{hu2021metasensing}, offering new ideas for enabling SSaC.
  \item \emph{THz techniques}: Exploiting the potential of higher frequencies say THz bands, will offer unique opportunities for sensing and environment reconstruction since it allow ultra-high resolution in all physical dimensions (including angle, range, Doppler). The current research trends of THz sensing techniques include THz imaging \cite{Li2021THzImaging}, THz massive MIMO \cite{wan2021THzMIMO}, and holographic MIMO systems.
\end{itemize}

\section{Conclusion}
By enabling a variety of applications and higher spectrum utilization, SSaC will be one of the key techniques to achieve the 6G vision. In this paper, we first present our understanding of the concept, vision, and evolution roadmap of SSaC. We then introduce three application scenarios of SSaC, followed by detailed description of several typical use cases, the corresponding performance metrics and key enabling technologies.

\section{Acknowledgment}
We would like to thank Ms. Ying Du, Dr. Xia Shen, Dr. Guochao Song, Mr. Yang Li from CAICT, Dr. Li Chen from USTC, and Dr. Peixi Liu from Peking University for their contribution to this work.

\bibliographystyle{ieeetr}

\end{document}

%% file: header.tex
\def\phi{\varphi}

\def\({\left(}
\def\){\right)}

\def\b0{{\mathbf{0}}}

\newcommand{\nn}{\nonumber}